\documentstyle[twocolumn,aps,pra,psfig]{revtex}

\begin{document}
\title{Quantum fluctuations for drag free geodesic motion}
\author{Jean-Michel Courty \thanks{courty@spectro.jussieu.fr} 
and Serge Reynaud \thanks{reynaud@spectro.jussieu.fr}}
\address{Laboratoire Kastler Brossel 
\thanks{Laboratoire de l'Universit\'{e} Pierre et Marie Curie,
 de l'Ecole Normale Sup\'{e}rieure 
et du Centre National de la RechercheScientifique},
 UPMC case 74, \\
4 place Jussieu, F-75252 Paris Cedex 05 }
\date{December 1999}
\maketitle

\begin{abstract}
The drag free technique is used to force a proof mass to follow a geodesic
motion. The mass is protected from perturbations by a cage, and the motion
of the latter is actively controlled to follow the motion of the proof mass.
We present a theoretical analysis of the effects of quantum fluctuations for
this technique. We show that a perfect drag free operation is in principle
possible at the quantum level, in spite of the back action exerted on the
mass by the position sensor.

{\bf PACS: 42.50 Lc; 04.80.Cc; 07.50-e}
\end{abstract}

\medskip

The Galilean principle of the universality of free fall has for a long time
had a limited accuracy due to technical difficulties. The effect of air drag
on falling bodies, the main of these difficulties, was already discussed at
length by Galileo and Newton \cite{Galileo,Newton}. Nowadays this effect may
be mastered either by an active control of the falling bodies or by letting
the fall take place in a vacuum drop tower. Using an active control greatly
reduces the vacuum requirement while allowing a test accuracy of $5\times
10^{-10}$ \cite{Faller87}. This accuracy does not reach the level of torsion
balance experiments \cite{Roll64,Braginsky72,Adelberger87} but it may do so
in the future with the additional use of a vacuum drop chamber \cite
{Kuroda89}. The accuracy of the measurement of relative acceleration of
freely falling bodies may thus reach $10^{-12}$ with a measurement time of $%
4.7$ s corresponding to a fall of $109$ m in the Bremen drop tower \cite
{Dittus96}.

The use of drag free technique has also been proposed for ultra high
accuracy satellite tests of the equivalence principle \cite{STEP96,LISA97}.
When such a technique is used, the freely falling proof mass is protected
from the non gravitational perturbations like residual air drag, radiation
pressure, {\it et cetera} by the satellite and its relative motion with
respect to the satellite is monitored by a high sensitivity position sensor.
The effect of these environmental perturbations on the satellite motion is
then finely compensated through the action of thrusters. In the present
paper, we evaluate the ultimate limits of the drag free technique. We study
the residual motion of a proof mass protected from environmental
perturbations by a cage, the motion of which is itself actively controlled
to follow the geodesic motion of the proof mass. These questions have
already been addressed from a classical point of view \cite{Jafry96,DeBra97}
which is sufficient for assessing the performance of real existing devices.

It is however very interesting to address the same questions in the context
of quantum measurement theory, at least as questions of principle. In a drag
free technique, the proof mass is continuously monitored by a position
sensor and it is therefore submitted to the back action of the sensor \cite
{Braginsky92}. When the sensibility of the sensor is improved, the back
action noise is expected to increase. It seems therefore difficult to have
at the same time a highly sensitive tracking and an unperturbed geodesic
motion of the proff mass. In fact, this difficulty may be circumvented by
the use of active techniques. We show in the present paper that drag free
operation may be in principle performed at the quantum level with the proof
mass following a nearly ideal geodesic trajectory. 

We consider in this paper a cold damped capacitive sensor developed at
ONERA\ for ultrasensitive accelerometry \cite
{Bernard91,Touboul92,Willemenot97}. The present analysis heavily relies on
previous studies of quantum and thermal fluctuations in such a device \cite
{Grassia98,Courty99,Grassia99}. In particular the analysis of the position
sensor will simply be taken from these references. The new result of the
present paper will concern the drag free system with the measured
acceleration between the proof mass and the cage used as an error signal to
servo control the cage motion. The drag free system is sketched on figure $1$%
.

The mechanical response of the system in the absence of servo control can be
written in term of a mechanical impedance matrix 
\begin{equation}
\left( 
\begin{array}{cc}
\Xi _{{\rm p}}+\Xi _{{\rm s}} & -\Xi _{{\rm s}} \\ 
-\Xi _{{\rm s}} & \Xi _{{\rm c}}+\Xi _{{\rm s}}
\end{array}
\right) \left( 
\begin{array}{c}
V_{{\rm p}} \\ 
V_{{\rm c}}
\end{array}
\right) =\left( 
\begin{array}{c}
F_{{\rm p}}^{{\rm t}} \\ 
F_{{\rm c}}^{{\rm t}}
\end{array}
\right)
\end{equation}
Throughout the paper the descriptions are given in the frequency domain and
the quantum convention is used for the Fourier transform. The electronics
convention may be recovered by substituting $j$ to $-i$. All physical
observables are represented as non commutative quantum operators.

The proof mass velocity $V_{{\rm p}}$ and the cage velocity $V_{{\rm c}}$
are determined through this equation to the total forces $F_{{\rm p}}^{{\rm t%
}}$ and $F_{{\rm c}}^{{\rm t}}$ acting on the two objects 
\begin{eqnarray}
F_{{\rm p}}^{{\rm t}} &=&F_{{\rm p}}+F_{{\rm s}}+F_{{\rm t}}  \nonumber \\
F_{{\rm c}}^{{\rm t}} &=&F_{{\rm c}}-F_{{\rm s}}-F_{{\rm t}}
\end{eqnarray}
Here $F_{{\rm p}}$ is external force acting on the proof mass despite the
screening (for example the gravitational force) and $F_{{\rm c}}$ is the
external force acting on the cage (including all the environmental
perturbations). $F_{{\rm s}}$ is the force exerted on the proof mass and on
the cage through the separating space between them (for example the Langevin
force associated with residual gas). $F_{{\rm t}}$ is the back action force
exerted by the electromechanical transducer used to measure the relative
position between the proof mass and the cage. The action and reaction forces
have been treated as equal, which amounts to neglect the delays at the low
mechanical frequencies considered here.

\begin{figure}[h]
\centerline{\psfig{figure=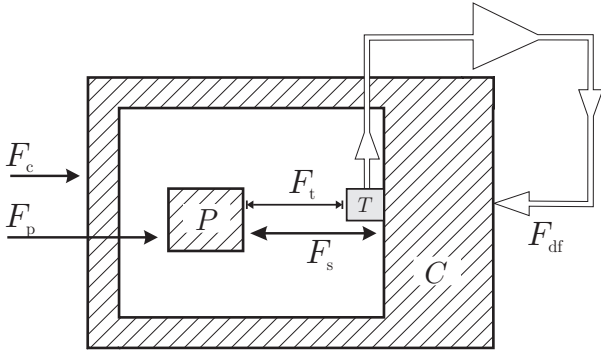,width=8cm}}
\caption{Simplified scheme of the drag free system. A proof
mass P is protected from environmental forces by a screening cage C.
The position of the proof mass relative to the cage is measured by a
capacitive transducer T used as a position sensor. Its output is used as an
error signal for servo controlling the motion of the cage. $F_{{\rm t}}$
and $F_{{\rm df}}$ represent respectively the back action force exerted by
the position sensor and the correcting action of the drag free system. $F_{%
{\rm c}}$ is the external force acting on the cage and $F_{{\rm p}}$ is the
external force acting on the proof mass despite the screening by the cage.
The separating space between the proof mass and the cage may also exert a
perturbation $F_{{\rm s}}$ onto the proof and the associated reaction onto
the cage.}
\label{figure1}
\end{figure}

$\Xi _{{\rm p}}$ is the mechanical impedance of the free proof mass, that is
a mass which would be neither perturbed by the environmental forces screened
by the cage nor disturbed by its coupling to a sensor. It takes in
particular into account the inertial mass $M_{{\rm p}}$ and the dissipative
effects associated with the unscreened forces. In this sense what we call in
the following the free motion of the proof mass is described by the equation 
\begin{equation}
\Xi _{{\rm p}}V_{{\rm p}}^{{\rm fr}}=F_{{\rm p}}  \label{massfree}
\end{equation}
When quantum fluctuations are taken into account (see for example \cite
{Reynaud92,Bocko96}) this equation is known to include the fluctuations
associated with Schrodinger equation at the limiting case of a vanishingly
small dissipation \cite{Jaekel93}. $\Xi _{{\rm c}}$ is similarly the
impedance of cage in the absence of coupling to the proof mass and to the
transducer. Finally $\Xi _{{\rm s}}$ corresponds to the interaction between
the proof mass and the cage through the restoring force, damping associated
with residual gases, back action forces due to the presence of the position
sensor {\it et cetera}.

The mechanical impedances and force fluctuations are related through the
fluctuation dissipation relations initially discovered by Einstein in his
analysis of Brownian motion \cite{Einstein05}. These relations are known as
Nyquist relations for electrical systems \cite{Nyquist28} and they are
written below with quantum fluctuations accounted for \cite{Callen51,Landau}%
. The dissipative part of the susceptibility functions is directly related
to the commutator of the quantum force fluctuations \cite{Jaekel93}. Then
force fluctuations $F$ are characterized by a noise spectrum $\sigma _{FF}$
with its well-known expression for a thermal equilibrium at a temperature 
\begin{eqnarray}
&&\left\langle F\left[ \omega \right] \cdot F\left[ \omega ^{\prime }\right]
\right\rangle =2\pi \ \delta \left( \omega +\omega ^{\prime }\right) \
\sigma _{FF}  \nonumber \\
&&\sigma _{FF}=2k_{B}\Theta 
\mathop{\rm Re}%
\Xi \left[ \omega \right]  \nonumber \\
&&k_{B}\Theta =\frac{\hbar \left| \omega \right| }{2}\coth \frac{\hbar
\left| \omega \right| }{2k_{B}T}  \label{thermal}
\end{eqnarray}
The symbol `$\cdot $' denotes a symmetrized product for quantum operators
(used to get rid of the ordering ambiguity), $k_{B}$ is the Boltzmann
constant and $T$ the temperature. We have introduced an effective
temperature $\Theta $ with $k_{B}\Theta $ representing exactly the energy
per mode of the fluctuations. This energy reproduces the classical result $%
k_{B}T$ at the high temperature limit and the zero point energy $\frac{\hbar
\left| \omega \right| }{2}$ at zero temperature.

The motion sensor is a capacitive transducer designed to detect the relative
acceleration between the cage and the proof mass. Its operation and
performances have been described in detail elsewhere \cite
{Bernard91,Touboul92,Willemenot97}. We recall here the properties of this
sensor when the various parameters are optimized \cite{Grassia98,Grassia99}.

The sensor measures the velocity difference $\delta V$ between the cage and
the proof mass 
\begin{equation}
\delta V=V_{{\rm c}}-V_{{\rm p}}
\end{equation}
Its performance can be discussed by introducing an estimator $\widehat{%
\delta V}$ representing the velocity $\delta V$ as it may be inferred from
the output of the sensor. After an appropriate scaling this estimator is the
sum of the velocity signal $\delta V$ and of a sensing error $V_{{\rm se}}$ 
\begin{equation}
\widehat{\delta V}=\delta V+V_{{\rm se}}
\end{equation}
The sensing error $V_{{\rm se}}$ and the back action force $F_{{\rm t}}$
exerted by the transducer onto the proof mass are the main parameters of
interest for evaluating the performance of the sensor. Here we use their
expressions taken from \cite{Grassia98,Grassia99} and discuss the drag free
system.

In the drag free system the measured velocity $\widehat{\delta V}$ is used
as an error signal to servo control the motion of the cage. It is worth
recalling that this error signal has been obtained after a preamplification
stage with a large gain \cite{Courty99}. As shown in \cite{Grassia99}, we
may greatly simplify the analysis of the whole system by considering the
further amplification stages as noiseless. This is true in particular for
the amplifications used in the feedback loops. Hence, we may write the servo
control force $F_{{\rm df}}$ acting on the cage as merely proportional to
the velocity estimator $\widehat{\delta V}$ 
\begin{equation}
F_{{\rm df}}=-G\ \widehat{\delta V}
\end{equation}
The factor $G$\ is the gain of the servo loop and it corresponds to an
effective mechanical impedance. With the feedback in operation, the
equations of motion are now written 
\begin{equation}
\left( 
\begin{array}{cc}
\Xi _{{\rm p}}+\Xi _{{\rm s}} & -\Xi _{{\rm s}} \\ 
-\Xi _{{\rm s}}-G & \Xi _{{\rm c}}+\Xi _{{\rm s}}+G
\end{array}
\right) \left( 
\begin{array}{c}
V_{{\rm p}} \\ 
V_{{\rm c}}
\end{array}
\right) =\left( 
\begin{array}{l}
F_{{\rm p}}^{{\rm t}} \\ 
F_{{\rm p}}^{{\rm t}}-GV_{{\rm se}}
\end{array}
\right) 
\end{equation}
In the limit of a large servo loop gain, the solution of these equations is
read 
\begin{eqnarray}
V_{{\rm p}} &=&\frac{1}{\Xi _{{\rm p}}}\left( F_{{\rm p}}+F_{{\rm s}}+F_{%
{\rm t}}-\Xi _{{\rm s}}V_{{\rm se}}\right)   \nonumber \\
V_{{\rm c}} &=&\frac{1}{\Xi _{{\rm p}}}\left( F_{{\rm p}}+F_{{\rm s}}+F_{%
{\rm t}}-\left( \Xi _{{\rm s}}+\Xi _{{\rm p}}\right) V_{{\rm se}}\right)  
\nonumber \\
&=&V_{{\rm p}}-V_{{\rm se}}  \label{Vdragfree}
\end{eqnarray}

These quantum equations have exactly the same form as in a classical
analysis of fluctuations. As a first result, this proves that the drag free
technique is also effective when quantum fluctuations are taken into
account. But it is more interesting to consider these equations as
describing the ultimate performance of the system as it would be limited by
quantum fluctuation processes and to address in this manner the questions
asked in the introduction.

As already explained, the free motion of the proof mass corresponds to the
unperturbed equation (\ref{massfree}). This free motion is recovered in (\ref
{Vdragfree}) superimposed to perturbations having various origins. $F_{{\rm s%
}}$ represents the Langevin forces associated with the dissipative part of
the impedance $\Xi _{{\rm s}}$ characterizing the coupling between the proof
mass and the cage. $F_{{\rm t}}$ is the back action exerted by the sensor on
the proof mass and $V_{{\rm se}}$ is the sensing error which appears
multiplied by the mechanical impedance $\Xi _{{\rm s}}$. If we concentrate
on questions of principle these terms can be arbitrarily reduced by
tailoring the impedances and the associated fluctuations. Quantum mechanics
does not prevent a perfect drag free operation.

In the following we focus our attention on a simple and realistic treatment
of the amplifier as a phase independent device characterized by an
equivalent noise temperature and noise impedance \cite{Courty99}. In this
situation, the back action noise and the sensing error can be written as 
\begin{eqnarray}
\sigma _{F_{{\rm t}}F_{{\rm t}}} &=&4k_{B}\Theta _{{\rm a}}\ \rho \left| \Xi
_{{\rm s}}\right|  \nonumber \\
\sigma _{V_{{\rm se}}V_{{\rm se}}} &=&4k_{B}\Theta _{{\rm a}}\frac{1}{\rho
\left| \Xi _{{\rm s}}\right| }  \nonumber \\
k_{B}\Theta _{{\rm a}} &=&\frac{\hbar \omega _{{\rm t}}}{2}\coth \frac{\hbar
\omega _{{\rm t}}}{2k_{B}T_{{\rm a}}}
\end{eqnarray}
They vary as conjugated noises as functions of the mechanical impedance $%
\left| \Xi _{{\rm s}}\right| $ and of the dimensionless ratio $\rho $ which
represents the impedance matching between the amplifier and the
electromechanical transducer \cite{Grassia99}. $\Theta _{{\rm a}}$ is the
effective temperature characterizing the preamplifier noise. It depends on a
temperature $T_{{\rm a}}$ and the frequency of operation $\omega _{{\rm t}}$
of the electrical detection circuit. The latter lies in the 100kHz range
that is at much higher frequencies than the mechanical frequencies $\Omega $
of interest. The correlation function of the proof mass velocity (\ref
{Vdragfree}) is thus 
\begin{eqnarray}
\left| \Xi _{{\rm p}}\right| ^{2}\sigma _{V_{{\rm p}}V_{{\rm p}}}
&=&2k_{B}\Theta _{{\rm p}} \mathop{\rm Re} \Xi _{{\rm p}}  \nonumber \\
&+&2k_{B}\Theta _{{\rm s}} \mathop{\rm Re} \Xi _{{\rm s}}+4k_{B}\Theta _{%
{\rm a}}\left| \Xi _{{\rm s}}\right| \left( \rho +\frac{1}{\rho }\right)
\label{sigma1}
\end{eqnarray}
The first line represents the free motion of the proof mass and the second
line the noises added respectively by the three additional terms in (\ref
{Vdragfree}). The noise added to the free motion of the test mass by the
drag free system has a minimum level when the impedances are matched so that 
$\rho =1$. The noise thus reaches its minimum value 
\begin{eqnarray}
\left| \Xi _{{\rm p}}\right| ^{2}\sigma _{V_{{\rm p}}V_{{\rm p}}}
&=&2k_{B}\Theta _{{\rm p}}\mathop{\rm Re} \Xi _{{\rm p}}  \nonumber \\
&+&2k_{B}\Theta _{{\rm s}}\mathop{\rm Re} \Xi _{{\rm s}}+8k_{B}\Theta _{{\rm %
a}}\left| \Xi _{{\rm s}}\right|  \label{sigmaVV}
\end{eqnarray}

Equation (\ref{sigmaVV}) describes in a quantitative manner the ultimate
performance of the drag free system as far as the residual motion of the
proof mass is concerned. The added noise contains the fluctuations $%
2k_{B}\Theta _{{\rm s}}\mathop{\rm Re} \Xi _{{\rm s}}$ corresponding to the
Langevin force associated with the dissipation between the proof mass and
the cage and a second term having the same order of magnitude when the
equivalent temperature $\Theta _{{\rm s}}$ and $\Theta _{{\rm a}}$ are
equal. In fact it may even be larger in this situation if the impedance $\Xi
_{{\rm s}}$ is mainly reactive. Note that this feature makes a difference
with the previously studied case where the accelerometer was used for
measuring the force acting on the proof mass \cite{Grassia99}. In the
previous case, the last term was reduced by a factor of the order of the
frequency transposition ratio $\frac{\Omega }{\omega _{{\rm t}}}$. Here in
contrast, we are concerned with the real motion of the proof mass and not
only with the accelerometry signal used as error signal. We do not benefit
of this frequency transposition ratio for the motion control. To illustrate
this point we rewrite the velocity noise (\ref{sigmaVV}) with the assumption
of a zero temperature that is to say a temperature small with respect to all
frequencies of interest 
\begin{equation}
\left| \Xi _{{\rm p}}\right| ^{2}\sigma _{V_{{\rm p}}V_{{\rm p}}}=\hbar
\Omega \mathop{\rm Re} \left( \Xi _{{\rm p}}+\Xi _{{\rm s}}\right) +4\hbar
\omega _{{\rm t}}\left| \Xi _{{\rm s}}\right|
\end{equation}
The frequency transposition gives a large weight to the noise added by the
servo control. The technique of frequency transposition so well adapted to
the measurement of a force with the accelerometer is of no help for a drag
free operation.

Coming back to the general case we may also write the velocity noise for the
actively controlled cage motion. The formula equivalent to (\ref{sigma1}) is
read as 
\begin{eqnarray}
\left| \Xi _{{\rm p}}\right| ^{2}\sigma _{V_{{\rm c}}V_{{\rm c}}}
&=&2k_{B}\Theta _{{\rm p}}\mathop{\rm Re}\Xi _{{\rm p}}+2k_{B}\Theta _{{\rm s%
}}\mathop{\rm Re}\Xi _{{\rm s}}  \nonumber \\
&+&4k_{B}\Theta _{{\rm a}}\left| \Xi _{{\rm s}}\right| \left( \rho +\frac{%
\left| \Xi _{{\rm p}}+\Xi _{{\rm s}}\right| ^{2}}{\rho \left| \Xi _{{\rm s}%
}\right| ^{2}}\right) 
\end{eqnarray}
It can be used to discuss the ultimate performance of the drag free system
when the emphasis is put on the geodesic motion of the cage. In this case a
different value of the impedance matching parameter $\rho $ has to be chosen
in order to optimize the performance of the system $\rho =\frac{\left| \Xi _{%
{\rm p}}+\Xi _{{\rm s}}\right| }{\left| \Xi _{{\rm s}}\right| }$and this
choice leads to a minimum velocity noise for the cage 
\begin{eqnarray}
\left| \Xi _{{\rm p}}\right| ^{2}\sigma _{V_{{\rm c}}V_{{\rm c}}}
&=&2k_{B}\Theta _{{\rm p}}\mathop{\rm Re}\Xi _{{\rm p}}  \nonumber \\
&+&2k_{B}\Theta _{{\rm s}}\mathop{\rm Re}\Xi _{{\rm s}}+8k_{B}\Theta _{{\rm a%
}}\left| \Xi _{{\rm p}}+\Xi _{{\rm s}}\right| 
\end{eqnarray}

Active techniques are now used in the operation of gravitational wave
detection with interferometers \cite{Richman98,Losurdo99}, in particular for
improving isolation of the mirors from the ground motion. The motion of the
first stage of the isolator is measured with accelerometers and compensated
with feedback action. In the last stage, servo control is also used to
perform the final positioning of the miror. The discussions presented in the
present paper could be applied to analyse the limits of these techniques.%
\newline

\noindent {\bf Acknowledgments}

\noindent We wish to thank Francesca Grassia, Pierre Touboul and Philippe
Tourrenc for stimulating discussions.

\end{document}